\documentstyle[amssymb,amsmath,aps,epsfig,float,twocolumn]{revtex}
\restylefloat{figure}
\newcommand{\e}{{\rm e}}

\title{Numerical study of quantum percolation}
\author{A.~Wei{\ss}e and H.~Fehske}
\address{Physikalisches Institut, Universit\"a{}t Bayreuth, 
  95440 Bayreuth, Germany\\
  {\rm (\today)}\\[0.5cm]}  
\address{~\parbox{14cm}{\rm
    We study the density of states and the optical conductivity of 
    the classical double-exchange model on a site percolated cluster.
    \vskip0.05cm\medskip PACS numbers: 
    }}
\begin{document}
\maketitle

In a recent attempt~\cite{wlf01b} to describe the metal-insulator
transition in CMR manganites we assumed a percolative coexistence of 
the two competing phases. In the insulating phase doped holes are trapped 
by local Jahn-Teller or breathing type lattice distortions. The metallic 
phase consists of itinerant carriers whose hopping amplitude is 
coupled via double-exchange~\cite{DE} to a background of localised 
spins. 

To gain insight into the properties of the latter phase, we study 
the classical double-exchange model on a site percolated cluster,
\begin{equation}\label{htb}
  H = \sum_{\langle kl\rangle}[ t_{kl}^{} c_{k}^{\dagger} c_{l}^{} 
  + \textrm{H.c.}]\,.
\end{equation}
Here $c_{l}^{\dagger}$ creates a spinless fermion in the Wannier 
state at site $l$, and the summation is over neighbouring sites 
on a simple cubic lattice.
The matrix element $t_{kl}^{}$ is nonzero only between sites of a 
cluster,
\begin{eqnarray}\label{tkl}
  t_{kl}^{} & = & \cos\tfrac{\theta_k}{2}
  \cos\tfrac{\theta_l}{2}\ \e^{-i(\phi_k-\phi_l)/2}
  \nonumber{}\\
  & & +\ \sin\tfrac{\theta_k}{2}
  \sin\tfrac{\theta_l}{2}\ \e^{i(\phi_k-\phi_l)/2}\,,
\end{eqnarray}
where the angles $\{\theta_l,\phi_l\}$ parameterise the background of 
classical spins. Hence, in the present model there are two types of 
disorder which cause scattering or localisation of the involved 
fermions, namely, the random structure of the cluster and the 
disorder in the hopping matrix elements.

To construct a certain realization of the model~(\ref{htb}) 
with a given probability $p$ we choose active sites from a simple 
cubic lattice. After labelling~\cite{HK76} the resulting clusters we
keep only the largest (or the spanning) one for further calculations.
To fix the hopping matrix elements a set $\{\theta_l,\phi_l\}$ is 
taken from an ensemble of thermalized classical spins in a magnetic 
field $h_{\rm eff}$. This introduces the parameter 
$\lambda=\beta g \mu_B h_{\rm eff}$. 

\begin{figure}[tb]
  \centering
  \epsfig{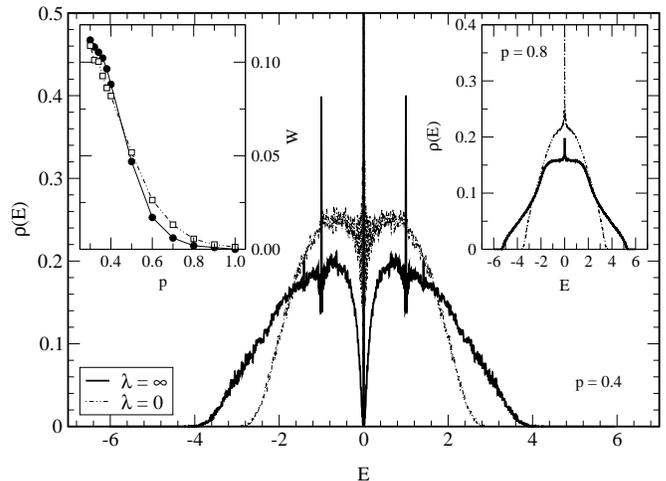}
  \caption{Density of states obtained for $p=0.4$ (right inset $p=0.8$) 
    and $\lambda=0,\,\infty$ on a $100^3$ site lattice with
    periodic boundary conditions; 
    Left inset: Weight of the central peak 
    as a function of $p$ for $\lambda=0,\,\infty$ and $64^3$ sites.
    }\label{figdos}
\end{figure}  

\begin{figure}[tb]
  \centering
  \epsfig{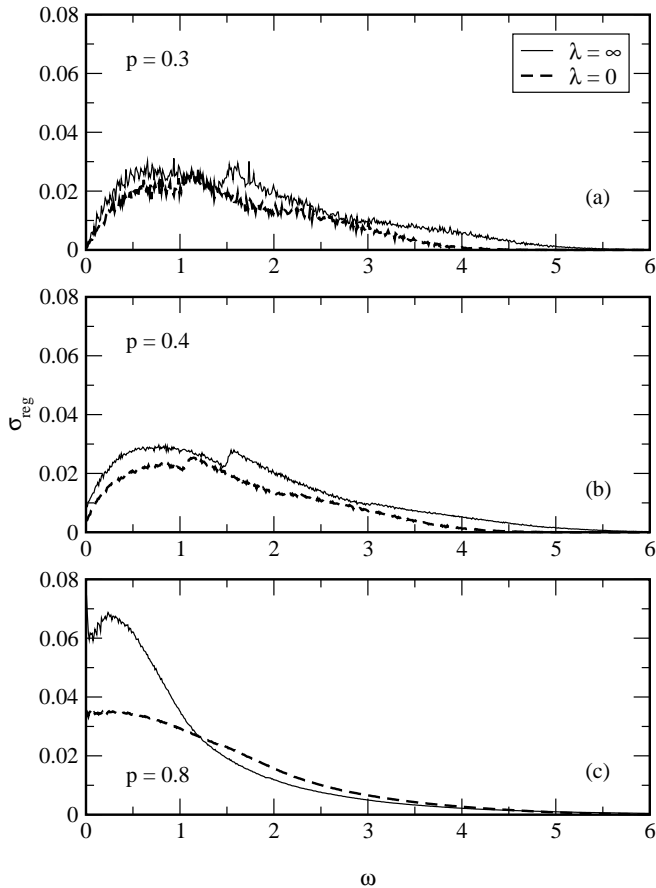}
  \caption{Optical conductivity for $p=0.3$, $0.4$, $0.8$ and
    $\lambda=0,\,\infty$ averaged over 700 realizations on a
    $10^3$ site lattice.}\label{figsig}
\end{figure}  

Application of Chebyshev expansion and maximum entropy methods~\cite{SR97}
yields the density of states, shown in Figure~\ref{figdos} for different 
values of $p$ and $\lambda$. 
For the ordered spin background ($\lambda = \infty$)
with decreasing $p$ a pseudo-gap feature appears close to the 
band centre, together with a distinct peak at $E=0$. 
The weight $W$ of this central peak (left hand inset) increases 
continuously, and makes up more than 10\% of the spectrum close to 
the classical percolation threshold $p_c \approx 0.3116$.
On the other hand, increasing spin disorder, i.e., decreasing $\lambda$, 
transfers spectral weight to the band centre. It does not
seem to affect the weight of the central peak.
Of course, both types of disorder reduce the overall bandwidth.

What remains to be clarified is the nature of the involved 
eigenstates. Our calculations indicate that the states 
in the band centre show a chequerboard structure and are
less localised compared to the rest of the spectrum. Similar 
behaviour was also found for lower dimensional systems, as 
discussed in Ref.~\cite{ITA94}.

Figure~\ref{figsig} shows the regular part of the optical
conductivity,
\begin{equation}
  \sigma_{\rm reg}(\omega) = \tfrac{1}{N} \sum_{m\neq 0} 
  \frac{|\langle m|j_x|0\rangle|^2}{\omega_{m}}\ 
  \delta(\omega - \omega_{m})\,, 
\end{equation} 
with $\omega_m=E_m-E_0$. The current $j_x$ is given by
\begin{equation}
  j_x = \sum_{\langle kl\rangle_x} [i\,t_{kl}^{} c_k^{\dagger} c_l^{}
  + \textrm{H.c.}]\,,
\end{equation}
where the summation extends over neighbouring sites in $x$-direction only.
Considering clusters on a $10^3$ site lattice, we calculated the 
eigenstates of the hopping matrix and summed up the current matrix 
elements between empty and occupied eigenstates. We assumed zero (electron) 
temperature and a band filling of $0.2$.
Clearly, below the classical percolation threshold (panel (a)) we deal 
with finite size clusters which do not (or rarely) connect the boundaries. 
Hence, the averaged $\sigma_{\rm reg}$ is rather noisy and approaches 
zero in the limit $\omega\rightarrow 0$. For $p>p_c$ (panels (b) and (c)) 
the curves become smooth, and we obtain a finite dc 
conductivity. In panel (b) the hump between $\omega=1$ and $2$ is due 
to excitations into the central peak. With decreasing $\lambda$ it is 
shifted to lower frequencies, because of the reduced bandwidth. 
Further studies of the model should clarify whether the quantum 
percolation threshold~\cite{KO99} $p_q > p_c$ is also visible in the
optical conductivity.

\end{document}